# Chondrocyte Heterogeneity; It Is the Time to Update the Understanding of Cartilage Histology

Yasser A. Ahmed

Department of Histology, Faculty of Veterinary Medicine, South Valley University, Qena, Egypt

Abstract

Chondrocytes were described as one cell populations in most cartilage literature. Two different chondrocyte populations; dark and light, were described in the articular cartilage and a third population, adipochondrocytes, was described in the elastic cartilage. The current literature of cartilage histology should be updated and highlight that three different populations of chondrocytes are existed in cartilage.

Keywords: Cartilage, Light chondrocytes, Dark chondrocytes, Adipochondrocytes. Chondrocyte populations.

Cartilage is present in the body in three major types; hyaline, fibrocartilage and elastic cartilage. All these types were reported to contain a single cell type, the chondrocyte, which is embedded within lacunae in the extracellular matrix. Hyaline cartilage is the most common type of cartilage and located on articular surfaces, in the respiratory tract and ribs. Elastic cartilage has an additional network of elastic fibers and is present in the ear pinnae. Fibrocartilage characterized by the unique arrangement of chondrocytes in parallel rows between bundles of collagen type I (Ahmed, 2007). Most of the literature describe chondrocytes as one cell population (Akkiraju and Nohe, 2015; Mescher, 2018). Interestingly, two populations of chondrocytes; dark and light were described in articular cartilage and growth plate (Fig.1) from many species in vivo and in vitro including; equine (Ahmed et al., 2007b), (Ahmed et al., 2007a), mice (Chen et al., 2010a; Chen et al., 2010b), quail and duck (Ahmed, 2012). The dark chondrocytes have electron-dense cytoplasm and dark nucleus (Fig. 1A) and the light chondrocytes characterized by electron-lucent cytoplasm and light nucleus (Fig. 1B). Both cells could be differentiated not only on the morphological bases but also on the gene expression level. Dark and light chondrocytes play different roles in building the cartilage matrix (Chen et al., 2010a). A third population of chondrocytes was mentioned in the elastic cartilage of some species and termed as lipochondrocytes in mice (Sanzone and Reith, 1976) and adipochondrocytes in rabbit (Ahmed and Abdelsabour-Khalaf, 2018). Both lipochondrocytes and adipochondrocytes have the same morphology (Fig. 2). These cells contain a large fat globule filling their cytoplasm and the nucleus is flattened and peripherally located giving the morphological appearance of the adipocytes. The physiological role of adipochondrocytes is to be illustrated. Thus, the literature, especially the medical books, should be updated and highlight that there are, at least, three different populations of chondrocytes; light, dark and adipocyte-like chondrocytes.

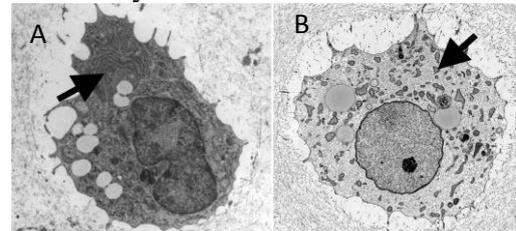

**Figure 1**: Dark (A) and light (B) chondrocytes with the transmission electron microscope. Arrows refer to RER (Ahmed, 2007).

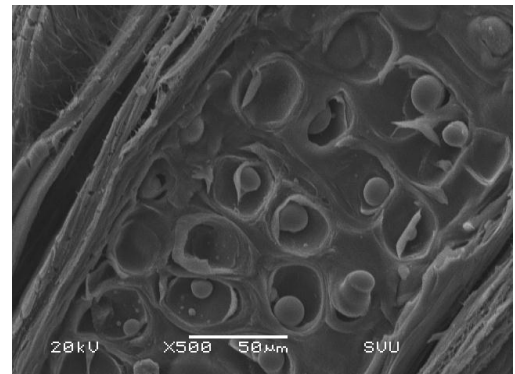

**Figure 2:** Adipochondrocytes with the scanning electron microscope (Ahmed and Abdelsabour-Khalaf, 2018).